\documentclass[aip,rsi,reprint,amsmath,amssymb,floatfix]{revtex4-2}
\usepackage[T1]{fontenc}
\usepackage[utf8]{inputenc}
\usepackage{graphicx}
\usepackage{bm}
\usepackage{lmodern}
\usepackage[hidelinks]{hyperref}
\begin{document}
\title{A Theoretical Investigation of Magnetic Susceptibility Measurement of Diamagnetic Liquids Using a Michelson Interferometer}
\author{David Shulman}
\email{davidshu@ariel.ac.il}
\affiliation{Department of Chemical Engineering, Ariel University, Ariel, Israel}
\affiliation{Physics Department, Ariel University, Ariel, Israel}
\date{November 22, 2026}
\begin{abstract}
We examine the theoretical use of a Michelson interferometer to infer the magnetic susceptibility of a diamagnetic liquid from a magnetically induced surface deformation. A ring magnet changes the liquid thickness along the sample beam, producing a signed phase shift relative to a fixed optical reference. For a double traversal of the sample, the fringe displacement is $m=2(n_\ell-n_a)\Delta z/\lambda$. We derive the susceptibility inversion both in a local gravity approximation and with surface tension included through the linearized Young--Laplace equation. The treatment distinguishes a finite reference field from a zero-field reference, specifies the displacement sign and susceptibility contrast, and identifies calibration and uncertainty requirements. Reproducible illustrative calculations use a uniformly magnetized annulus. The results establish a theoretical measurement relation and its assumptions; they do not constitute an experimental demonstration or a verified accuracy claim.
\end{abstract}
\maketitle

\section{Introduction}
Magnetic susceptibility describes the linear response $\mathbf M=\chi\mathbf H$ of a weakly magnetic material. Here $\chi$ denotes dimensionless SI volume susceptibility, not mass susceptibility. Force measurements with a Faraday balance probe the response to a magnetic-field gradient \cite{morris1968faraday}; this method must be distinguished from magneto-optical Faraday rotation. Optical measurement of a field-induced liquid-surface deformation provides another route to susceptibility. The Moses-effect method has been used to infer susceptibility from the surface slope while accounting for surface tension \cite{SHULMAN2023}.

We investigate a Michelson interferometer as a phase-sensitive readout of the same magnetically driven surface deformation. The optical observable is a change in optical path, from which susceptibility is inferred through a mechanical and magnetic model. This is a theoretical feasibility study: the numerical examples below are model predictions, not measurements. No experimental accuracy or superiority over existing techniques is asserted.

\section{Optical geometry and phase measurement}
\begin{figure}[t]
\centering
\includegraphics[width=\linewidth]{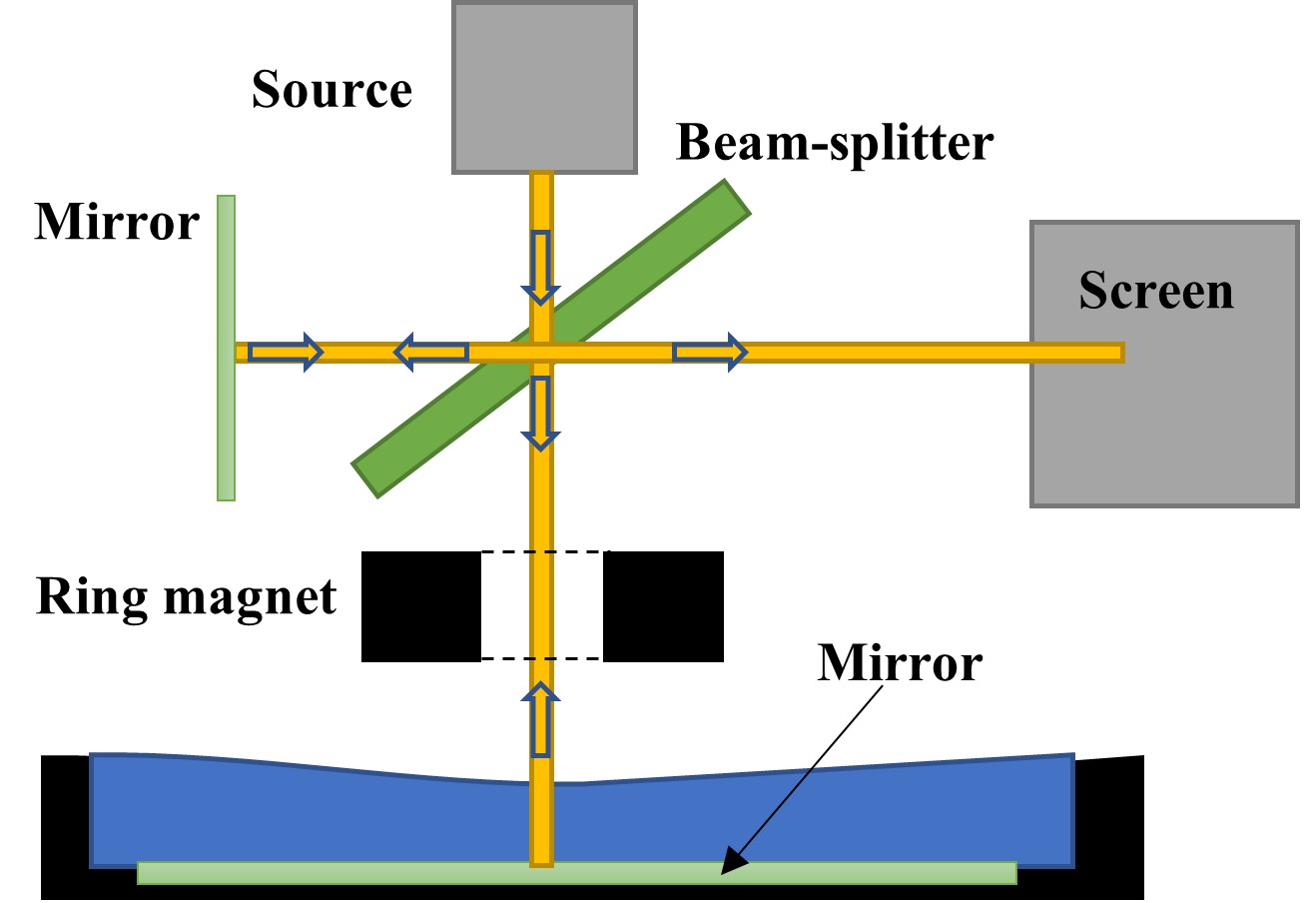}
\caption{Michelson geometry with a fixed submerged sample mirror and a double traversal of the liquid. The drawing is schematic and not to scale; only the magnet moves between measurements.}
\label{fig:setup}
\end{figure}

The sample arm passes through the bore of a ring magnet, crosses the liquid surface, reflects from a fixed mirror beneath the liquid, and retraces its path. A fixed mirror terminates the reference arm. The beams recombine at the same beam splitter (Fig.~\ref{fig:setup}). The sample mirror does not follow the liquid surface: the measured phase shift arises from replacement of gas path by liquid path on the outward and return traversals. Mirror motion would add a separate displacement signal.

Let $z(r,h)$ be upward surface displacement from the undisturbed level, where $r$ is radius and $h$ is the gap between the lower magnet face and the undisturbed surface. A depression has $z<0$. Let $\Delta z=z(r,h_1)-z(r,h_0)$ and $\Delta n=n_\ell-n_a$, with $n_\ell$ and $n_a$ the liquid and ambient-gas refractive indices at the vacuum wavelength $\lambda$. Fixed optical endpoints imply that an increase in liquid path replaces an equal length of gas path. For $q$ traversals of the moving interface,
\begin{align}
\Delta L&=q\Delta n\,\Delta z,\\
\Delta\phi&=\frac{2\pi q\Delta n}{\lambda}\Delta z,\\
m\equiv\frac{\Delta\phi}{2\pi}&=\frac{q\Delta n}{\lambda}\Delta z.
\label{eq:fringes}
\end{align}
In this paper $q=2$. The phase shift and fringe displacement $m$ are signed; counting fringes without their direction yields only a magnitude. A change of phase origin or detector port changes the measured sign convention, which must be calibrated. Fractional fringes can be recovered using phase stepping or quadrature detection and phase unwrapping.

For coherent beams of intensities $I_s$ and $I_r$, the detector signal can be written
\begin{equation}
I=I_s+I_r+2V\sqrt{I_sI_r}\cos\phi,
\end{equation}
where $0\leq V\leq1$ describes coherence and mode overlap. A single intensity reading does not determine a unique signed phase. Destructive interference gives zero intensity only for equal beam intensities and unit visibility. The analysis assumes near-normal incidence, negligible beam steering, stable indices, and a beam narrow enough to sample a nearly uniform surface height. For a broad beam, the detected complex field must instead be integrated over the transverse phase distribution.

\section{Surface deformation and susceptibility}
\subsection{Assumptions and sign convention}
Define the susceptibility contrast $\chi_c=\chi_\ell-\chi_a$ and density difference $\Delta\rho=\rho_\ell-\rho_a$. The leading-order weak-susceptibility approximation uses the externally imposed field $B$ without its small perturbation by the liquid. The model assumes a static, incompressible liquid with constant surface tension $\gamma$, a large bath, small slopes, and a deformation small relative to the field-variation length. The field is evaluated at the undeformed interface. For a finite bath, the wall condition and volume conservation must be included; the infinite-bath solution below is then only an approximation.

An energy functional, apart from a constant, is
\begin{equation}
\mathcal E=\int\!\left[\frac{\Delta\rho g}{2}z^2+
\frac{\gamma}{2}|\nabla z|^2-\frac{\chi_c B^2}{2\mu_0}z\right]dA.
\end{equation}
Its stationary point gives the linearized axisymmetric Young--Laplace balance
\begin{equation}
\Delta\rho g z-\gamma\left(z''+\frac{z'}r\right)
=\frac{\chi_c B^2(r,h)}{2\mu_0}.
\label{eq:balance}
\end{equation}
The conditions are $z'(0)=0$ and $z(r)\to0$ as $r\to\infty$. Equation~(\ref{eq:balance}) is a small-slope equation, not the full nonlinear curvature equation. With this upward-positive convention, $\chi_c<0$ produces a depression. The capillary length is $\ell_c=\sqrt{\gamma/(\Delta\rho g)}$.

If the field varies on a lateral scale much larger than $\ell_c$, surface tension can be neglected locally:
\begin{equation}
z(r,h)\simeq\frac{\chi_c B^2(r,h)}{2\mu_0\Delta\rho g}.
\label{eq:local}
\end{equation}
This approximation is useful for scaling, but is not automatically valid near the bore or edge of a ring magnet. The density and susceptibility of air may be neglected only when their contributions are below the required uncertainty. In particular, the measured contrast must be converted back using $\chi_\ell=\chi_c+\chi_a$.

\subsection{Local and capillary-corrected inversion}
Let $D=B^2(r,h_1)-B^2(r,h_0)$. Combining Eqs.~(\ref{eq:fringes}) and (\ref{eq:local}) gives
\begin{equation}
\boxed{\chi_c=\frac{2\mu_0\Delta\rho g\lambda m}{q\Delta n D}.}
\label{eq:inversion}
\end{equation}
For the present Michelson arrangement, this becomes
\begin{equation}
\chi_c=\frac{\mu_0\Delta\rho g\lambda m}{\Delta n D}.
\end{equation}
For motion toward a stronger field, $D>0$ and a diamagnetic depression gives $m<0$ under the stated convention. Division by a small $D$ is ill-conditioned.

For the capillary-corrected calculation, define $K(r,h)$ as the solution of
\begin{equation}
\Delta\rho g K-\gamma\left(K''+\frac{K'}r\right)
=\frac{B^2(r,h)}{2\mu_0}
\label{eq:kernel}
\end{equation}
with the same boundary conditions. Linearity gives $z=\chi_c K$ and
\begin{equation}
\boxed{\chi_c=\frac{\lambda m}{q\Delta n\,[K(r,h_1)-K(r,h_0)]}.}
\label{eq:fullinversion}
\end{equation}
The difference of two responses is essential. A formula for the absolute height cannot be applied directly to a measured height change unless the reference field and deformation are negligible. The Green-function form of $K$ is given in Appendix~\ref{app:kernel}.

For measurements $m_i$ at several radii or magnet positions, define $A_i=q\Delta n\Delta K_i/\lambda$. With independent phase variances $\sigma_i^2$, a weighted estimate is
\begin{equation}
\widehat\chi_c=\frac{\sum_i A_i m_i/\sigma_i^2}{\sum_i A_i^2/\sigma_i^2}.
\end{equation}
This expression treats the model coefficients as known. Uncertain magnetic calibration, correlations, background offsets, and drift require a joint fit or covariance propagation; adding more phase samples does not remove shared systematic errors.

\section{Illustrative calculations}
\subsection{Optical scale}
As a prescribed displacement example, take $|\Delta z|=70\,\mu\mathrm m$, $\lambda=633\,\mathrm{nm}$, $n_\ell=1.333$, and $n_a=1.0003$. Equation~(\ref{eq:fringes}) gives $|m|=73.58$ for $q=2$. This example is not a prediction that a particular magnet produces $70\,\mu\mathrm m$ at a specified gap. Both liquid and gas path changes have been included. For the same displacement, a Michelson double pass produces twice the phase shift of a single-pass Mach--Zehnder arrangement; this comparison assumes equal refractive indices and wavelength, not equal realized measurement noise.

\subsection{Ring-magnet model}
To make the numerical examples reproducible, we model a uniformly axially magnetized annulus of length $L=30\,\mathrm{mm}$, inner radius $a_i=2.5\,\mathrm{mm}$, and outer radius $a_o=11.5\,\mathrm{mm}$. Its magnetization is an illustrative input $M=8.0\times10^5\,\mathrm{A\,m^{-1}}$, not a fitted or measured property of the original apparatus. The field is the superposition of the outer and oppositely directed inner cylindrical surface currents, as described in Appendix~\ref{app:field}. A single solid-solenoid field does not represent the bore.

The illustrative liquid parameters are $\Delta\rho=998\,\mathrm{kg\,m^{-3}}$, $\gamma=0.072\,\mathrm{N\,m^{-1}}$, $\chi_c=-9.05\times10^{-6}$, and $g=9.81\,\mathrm{m\,s^{-2}}$, giving $\ell_c\simeq2.71\,\mathrm{mm}$. These are water-like model inputs; ambient-gas density and susceptibility are set to zero for this example. They must be replaced by temperature-dependent sample and ambient values in a measurement. We use $\mu_0\simeq4\pi\times10^{-7}\,\mathrm{H\,m^{-1}}$ to the precision needed here.

\begin{figure*}[t]
\centering\includegraphics[width=.95\textwidth]{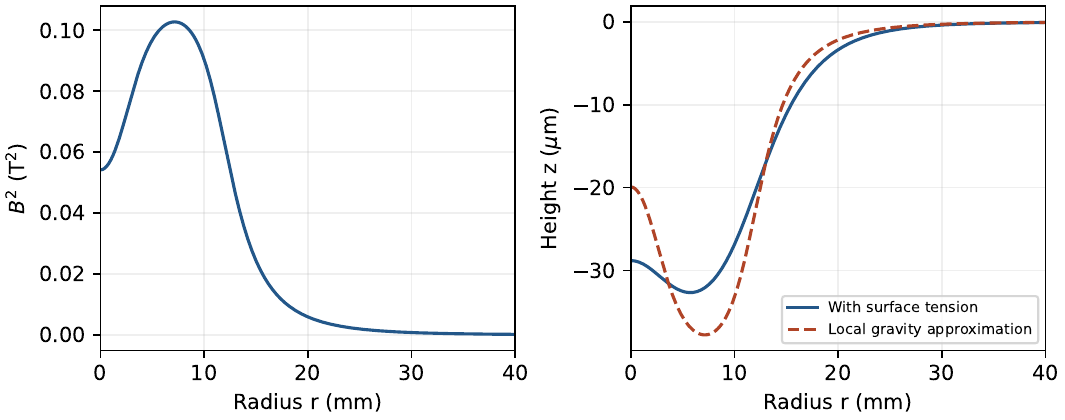}
\caption{Illustrative ring-magnet calculation at $h=3\,\mathrm{mm}$ using the stated parameters. Left: squared field magnitude on the undeformed interface. Right: upward-positive surface height from the linear capillary model and the local gravity approximation. These are synthetic predictions, not experimental data.}
\label{fig:profile}
\end{figure*}
\begin{figure*}[t]
\centering\includegraphics[width=.95\textwidth]{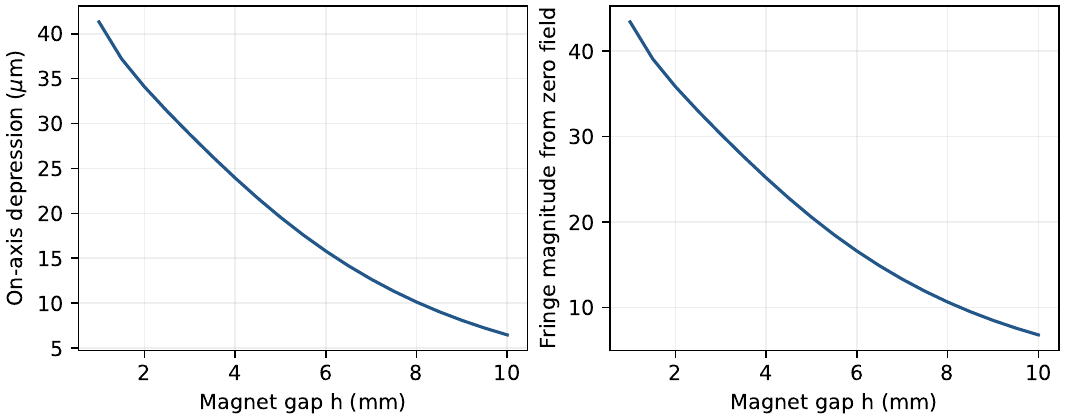}
\caption{Predicted on-axis depression and fringe magnitude relative to a zero-field reference for the illustrative parameters. The optical path samples $r=0$, which need not coincide with the deepest point of an annularly forced surface. A finite reference gap requires subtraction of its response.}
\label{fig:response}
\end{figure*}

The field is evaluated by Gauss--Legendre quadrature and Eq.~(\ref{eq:balance}) by a radial finite-difference solve with regularity at the axis and $z(R)=0$ at $R=120\,\mathrm{mm}$. The radial spacing is $0.2\,\mathrm{mm}$. At a gap of $3\,\mathrm{mm}$, the predicted on-axis height is approximately $-28.81\,\mu\mathrm m$, with maximum radial slope $0.0035$. These slopes support the small-slope approximation for this example, but do not establish the absence of other model errors.

Numerical checks compare the field on the axis with its analytic expression, double the field-quadrature order from 64 to 128 at a $1\,\mathrm{mm}$ gap, halve the radial spacing at a $3\,\mathrm{mm}$ gap, and extend the radial domain to $180\,\mathrm{mm}$. Grid refinement changes the on-axis height by less than $0.02\%$ for the $3\,\mathrm{mm}$ case. A manufactured Gaussian solution independently checks the radial operator and sign. These are discretization checks at the specified conditions, not a bound on physical-model uncertainty or a claim of experimental accuracy. The scripts and numerical summaries accompany the manuscript.

\section{Proposed measurement and uncertainty}
\subsection{Measurement protocol}
Align the probe close to the magnet axis and normal to the liquid surface, keeping all optical endpoints mechanically fixed. Map the vector magnetic field at the relevant gaps or validate a magnet model against a calibrated field probe. Record the liquid temperature, density, surface tension, refractive index, and ambient conditions. Translate the magnet between known positions, wait for hydrodynamic relaxation, and acquire a time-resolved phase sequence with a recorded sign convention. Use repeat forward and reverse scans and a stationary-magnet control to identify drift, vibration, evaporation, or hysteresis in the mechanics. Record beam position and waist, since averaging and misalignment can change the inferred response.

A liquid with known susceptibility provides a useful calibration check. If a known field shape is written $B=s\widetilde B$, then the signal is proportional to $\chi_c s^2$. One unknown liquid cannot identify both $\chi_c$ and $s$ from phase alone. Independent field calibration or a reference liquid is necessary. Reference and sample liquids generally have different $\Delta\rho$, $\gamma$, and $\Delta n$, so their response kernels must be recomputed; a fringe ratio alone is insufficient. Calibration does not prove that all systematic errors have been removed.

\subsection{Uncertainty propagation}
Write $\sigma_x$ for a standard uncertainty, reserving $\Delta$ for differences between measurement states. For known $\Delta n$ and $\lambda$, the height uncertainty caused by phase noise is
\begin{equation}
\sigma_z=\frac{\lambda}{q|\Delta n|}\sigma_m
=\frac{\lambda}{2\pi q|\Delta n|}\sigma_\phi.
\end{equation}
For the local inversion, an independent-input first-order estimate, neglecting uncertainty in $\mu_0$, is
\begin{align}
\left(\frac{\sigma_{\chi_c}}{|\chi_c|}\right)^2
&\simeq\left(\frac{\sigma_m}{|m|}\right)^2
+\left(\frac{\sigma_{\Delta n}}{|\Delta n|}\right)^2
+\left(\frac{\sigma_\lambda}{\lambda}\right)^2\nonumber\\
&\quad+\left(\frac{\sigma_{\Delta\rho}}{\Delta\rho}\right)^2
+\left(\frac{\sigma_g}{g}\right)^2
+\left(\frac{\sigma_D}{|D|}\right)^2.
\label{eq:uncertainty}
\end{align}
For $F_j=B^2(r,h_j)$, $D=F_1-F_0$. Gap uncertainty contributes
\begin{align}
\sigma_{D,h}^2={}&(\partial_hF_1)^2\sigma_{h_1}^2
+(\partial_hF_0)^2\sigma_{h_0}^2\nonumber\\
&-2(\partial_hF_1)(\partial_hF_0)\operatorname{Cov}(h_1,h_0).
\end{align}
The two gaps may share a common zero error, so treating them as independent is not always justified. A common field scale $s$ gives a relative susceptibility-scale uncertainty of $2\sigma_s/|s|$ to first order.

For the capillary inversion, use the Jacobian of Eq.~(\ref{eq:fullinversion}), including the dependence of $\Delta K$ on $\gamma$, $\Delta\rho$, geometry, position, and field parameters. In general $\sigma_{\chi_c}^2=J\Sigma J^{\mathsf T}$, with covariance matrix $\Sigma$; Monte Carlo propagation is preferable near a small response difference or with nonlinear parameter uncertainty. Density and gravity must not also be added as independent prefactors when already propagated through $K$. Model discrepancies, such as finite-bath effects, nonuniform magnetization, field evaluation at a displaced interface, and optical beam averaging, require separate assessment. Thus a gap tolerance alone cannot establish an overall percentage accuracy.

\section{Discussion and conclusion}
The Michelson double pass doubles phase sensitivity to a given height change relative to a single-pass Mach--Zehnder arrangement. For equal phase noise this halves the phase-derived height uncertainty. It does not reduce uncertainties in field calibration, surface tension, density, or magnet position by the same factor. The submerged mirror requires chemical compatibility and mechanical stability, and the return beam must maintain sufficient spatial overlap.
Both interferometric variants probe the Moses-effect surface response. Phase readout measures an optical-path change rather than a local surface slope, but it remains sensitive to surface refraction, alignment, environmental drift, and the magnetic calibration. Neither a larger magnet diameter nor a larger number of fringes guarantees a smaller total uncertainty. The geometry, capillary smoothing, and optical access must be considered together.

The proposed Michelson arrangement provides a consistent relation between signed fringe displacement and susceptibility contrast, including a finite reference-field correction and surface tension within linear theory. The illustrative calculations demonstrate how to evaluate this relation reproducibly. Experimental validation, a measured field map, and a complete uncertainty budget are still required before an accuracy claim can be made. The closely related Michelson and Mach--Zehnder studies share the same mechanical response; their primary distinction is optical geometry and pass number, not two independently validated susceptibility methods.

\appendix
\section{Green function for the linear surface model}\label{app:kernel}
Let $f(r,h)=B^2(r,h)/(2\mu_0\gamma)$. For an infinite axisymmetric bath the kernel in Eq.~(\ref{eq:kernel}) is
\begin{align}
K(r,h)={}&K_0(r/\ell_c)\int_0^r f(s,h)I_0(s/\ell_c)s\,ds\nonumber\\
&+I_0(r/\ell_c)\int_r^\infty f(s,h)K_0(s/\ell_c)s\,ds.
\end{align}
Here $I_0$ and $K_0$ are modified Bessel functions. At $r=0$, take the finite limiting value: the first term vanishes and the second integral remains. The positive kernel gives $z=\chi_c K<0$ for negative contrast. It approaches Eq.~(\ref{eq:local}) as capillary effects become negligible. The solution is exact for the stated linear equation and boundary conditions, not for arbitrary slopes or container geometries. In a closed finite bath, a pressure offset enforcing volume conservation and the actual contact-line condition replace the infinite-bath boundary condition.

\section{Magnetic field of a uniformly magnetized annulus}\label{app:field}
The surface-current representation of a uniformly magnetized cylinder is equivalent to an ideal solenoid \cite{derby2010,Shulman2023Mag}. For an annulus, the outer cylindrical sheet carries azimuthal current density $+M$ and the inner sheet $-M$. Place the observation plane at vertical coordinate zero, and let the magnet occupy heights $u\in[h,h+L]$. For a sheet of radius $a$ with positive current density $M$, define
\begin{equation}
Q=r^2+a^2-2ar\cos\varphi+u^2.
\end{equation}
Direct Biot--Savart integration gives
\begin{align}
\mathcal B_r(a)&=-\frac{\mu_0Ma}{4\pi}\int_h^{h+L}\!du
\int_0^{2\pi}\!\frac{u\cos\varphi}{Q^{3/2}}\,d\varphi,\\
\mathcal B_z(a)&=\frac{\mu_0Ma}{4\pi}\int_h^{h+L}\!du
\int_0^{2\pi}\!\frac{a-r\cos\varphi}{Q^{3/2}}\,d\varphi.
\end{align}
The ring field is $B_j=\mathcal B_j(a_o)-\mathcal B_j(a_i)$ for $j=r,z$, and $B^2=B_r^2+B_z^2$. Reversing the magnetization reverses both components but not $B^2$. On the axis $\mathcal B_r=0$ and
\begin{equation}
\mathcal B_z(a)=\frac{\mu_0M}{2}\left[
\frac{h+L}{\sqrt{(h+L)^2+a^2}}-
\frac{h}{\sqrt{h^2+a^2}}\right].
\end{equation}
This expression provides an independent check on the quadrature. Real permanent magnets may require a measured field map or a model including nonuniform magnetization; nominal dimensions alone do not fix field amplitude.

\section*{Source availability}
The manuscript sources, figures, and Python code for both interferometer studies are available at \url{https://github.com/shulm/Interferometry}. Numerical figures are generated from the explicitly stated illustrative parameters; no experimental dataset is reported.
\section*{Conflict of interest}
The author has no conflicts to disclose.
\bibliographystyle{apsrev4-2}
\bibliography{references}

@article{morris1968faraday,
  title={Faraday balance for measuring magnetic susceptibility},
  author={Morris, Bernard L and Wold, Aaron},
  journal={Review of Scientific Instruments},
  volume={39},
  number={12},
  pages={1937--1941},
  year={1968},
  publisher={American Institute of Physics}
}

@article{Shulman2023Mag,
  author = {Shulman, D.},
  title = {Semi-Analytical and Experimental Investigation of the Magnetic Field Generated by a Permanent Magnet},
  journal = {J. Magn.},
  volume = {28},
  number = {1},
  pages = {64--71},
  year = {2023},
doi={10.4283/JMAG.2023.28.1.064}
}

@article{SHULMAN2023,
title = {Measurement of magnetic susceptibility of diamagnetic liquids exploiting the Moses effect},
journal = {Journal of Magnetism and Magnetic Materials},
pages = {170553},
year = {2023},
issn = {0304-8853},
doi = {10.1016/j.jmmm.2023.170553},
author = {D. Shulman and M. Lewkowicz and E. Bormashenko},
}

@article{derby2010,
 author={Derby, Norman and Olbert, Stanislaw},
 title={Cylindrical magnets and ideal solenoids},
 journal={American Journal of Physics},
 volume={78}, pages={229--235}, year={2010},
 doi={10.1119/1.3256157}
}
\end{document}